\documentclass[twocolumn,showpacs,amssymb,amsmath,nobibnotes,prl]{revtex4}

\usepackage{graphicx}
\usepackage{xcolor}
\usepackage{bm}%

\begin{document}
\title{Explosive Instability in Keplerian Disks}
\author{Yuri Shtemler} \email{shtemler@bgu.ac.il}
\author{Edward Liverts} \email{eliverts@bgu.ac.il} \author{Michael Mond} \email{mond@bgu.ac.il}\affiliation{Department of Mechanical
Engineering, Ben-Gurion University of the Negev, Beer-Sheva 84105,
Israel}
\date{\today}

\begin{abstract}
In this paper it is shown that differentially rotating disks that are in the presence of weak axial magnetic field are prone to a new nonlinear explosive instability. The latter occurs due to the near-resonance three-wave interactions of a magnetorotational instability with stable Alfv\'{e}n-Coriolis and magnetosonic modes. The dynamical equations that govern the temporal evolution of the amplitudes of the three interacting modes are derived. Numerical solutions of the dynamical equations indicate that small frequency mismatch gives rise to two types of behaviour: 1. explosive instability which leads to infinite values of the three amplitudes within a finite time, and 2. bounded irregular oscillations of all three amplitudes. Asymptotic solutions of the dynamical equations are obtained for the explosive instability regimes and are shown to match the numerical solutions near the explosion time.
\end{abstract}

\maketitle

\textit{Introduction - }
The magnetorotational instability (MRI, \cite{Velikho1959} \nocite{Chandra1959}  -- {\cite{BALBUS:1991sp}) is believed to play a key role in the angular momentum transfer in accretion disks, and has been thoroughly investigated through linear analysis as well as nonlinear magnetohydrodynamic (MHD) simulations under a wide range of conditions and applications. First attempts to achieve analytical insight into the nonlinear evolution of the MRI focused on the dissipative saturation of the instability (\cite{Knobloch:2005la}, \cite{Umurhan:2007eu}) in environments that are characteristic of laboratory experiments. Recently it has been suggested that the mechanism of dissipationless wave interaction plays an important role in the nonlinear development of the MRI in astrophysical disks (\cite{Liverts:2012fr}, \cite{Shtemler:2013qv}).  According to one such scenario the MRI forms a triad of interacting modes with a stable slow or fast Alfv{\'e}n-Coriolis (AC) mode and a stable magnetosonic (MS) mode. This is a generalisation and adaptation to the thin, rotating, axially stratified disks of the well known three-wave interaction in static, homogeneous and infinite plasmas (\cite{Galeev:1961}, \cite{SagdeevandGaleev}). In previous works the saturation of the MRI by non-resonant excitation of a MS wave, as well as by resonantly exciting two small-amplitude linearly stable modes have been investigated (\cite{Liverts:2012fr}, \cite{Shtemler:2013qv}, \cite{Shtemler:2011bh}). It has been shown that in the resonant case the two linearly stable modes grow exponentially due to the nonlinear coupling to the saturated MRI. The effect of a small mismatch between the eigen-frequencies of the three modes is investigated in the current work. The presence of a small frequency mismatch does not merely lead to small quantitative deviations from the strict-resonance case, but may change the behaviour of the system in a fundamental way \cite{WERSINGER:1980fh}. In particular it is shown that near-resonance interaction may give rise to an explosive instability under which the amplitudes of the three interacting modes reach infinite values within a finite time. The explosion instability is a well known phenomenon in infinite homogeneous plasmas (\cite{Dikasov} \nocite{DUM:1969qy}\nocite{rosenbluthetal}  \nocite{Rosenbluth} \nocite{Weiland} -- \cite{Craik}). Indeed, Zakharov and Manakov \cite{ZAKHAROV:1973bu} divide the general resonance interactions of three one-dimensional wave packets in a non dissipative medium into parametric decay instabilities and explosive instabilities. While the former was investigated in \cite{Shtemler:2013qv}, the current research focuses on the explosive instability.

\textit{The Physical Model - } The thin disk asymptotic expansion procedure (\cite{REGEV:1983pb},
\cite{Shtemler:2007ly}) is applied to the magnetohydrodynamic (MHD) equations in order to study the weakly nonlinear evolution of the MRI in Keplerian disks. A detailed description of that procedure and its results for the steady-state as well as the linear problem is presented in \cite{Shtemler:2011bh}. The main results are hereby summarized:
1. Steady-State: Assuming axially isothermal steady-state the normalized mass density profiles are given by $n(r,\zeta)=N(r)\Sigma(\eta)$, where $\Sigma(\eta)=e^{-\eta ^2/2}$, $N(r)$ is an arbitrary function of $r$, the radial coordinate, $\eta =\zeta/H(r)$, $\zeta=z/\epsilon$ is the stretched axial coordinate, and $H(r)$ is the semi thickness of the disk. The latter [or alternatively the temperature profile $T(r)$] is an arbitrary function of $r$.
2. Linear perturbations: Modifying the axial mass density profile to ${\bar\Sigma} (\eta) =\mathrm{sech} ^2\eta$ enables the analytical solution of the linearized set of equations for small perturbations. The resulting eigenmodes are thus divided into two families. The first family, the Alfv\'{e}n-Coriolis (AC), represents in-plain perturbations and form a discrete spectrum whose eigenfunctions $\omega _{k,l}$ are labeled by two integers $-\infty <k<\infty,  \;l=\pm1$ such the $l=-1(+1)$ represents the slow (fast) AC modes and $k$ is the axial mode number. The fast AC modes are stable while the slow AC modes may become unstable. The number of unstable slow AC modes is determined by the local plasma beta which is given by $\beta(r)=\beta_0 N(r)C_s^2(r)/B_z^2(r)$ where $\beta_0$ is the beta value at some reference radius, and $C_s(r)=H(r)\Omega(r)$ where $\Omega(r) \propto r^{-3/2}$ is the Keplerian rotation velocity, and $B_z(r)$ are some arbitrary profiles of the sound velocity and the axial steady-state magnetic field, respectively. Thus, the threshold for exciting $k$ unstable modes is given by $\beta _{cr}^k=k(k+1)/3, k=1,2,\ldots$. It is those unstable slow AC modes that constitute the MRI whose eigenvalues are given by $\omega _{m,-1}=i\gamma _m, \;m=1, \dots,k$. Of particular importance is the fact that for $\beta=\beta _{cr}^k$, $\gamma _k=0$ is a double root of the dispersion equation. The eigenfunctions of both sets of AC modes may be expressed in terms of the Legendre polynomials.   The other family of eigen-oscillations in thin Keplerian disks includes the vertical magnetosonic (MS) modes. The latters are stable, possess a continuous spectrum, and their eigenfunctions are localized about the mid-plain and may be expressed in terms of some special functions \cite{Liverts:2012fr}. The two families of the linear eigenmodes, namely the AC and the MS modes, are the building blocks of the nonlinear analysis to be unfolded in the next sections.

\textit{Near-resonance interactions -- } The scenario that is introduced in the current work is a generalisation of the mechanism described in \cite{Shtemler:2013qv}: A $\beta$ value slightly above the first threshold for MRI is considered. As a results there is only one unstable MRI mode, that is characterised by axial wave number $k=1$ and whose growth rate is denoted by $\gamma$. A large amplitude MRI eigenmode (characterised by frequency $0+i\gamma $) forms a triad of interacting modes with a stable fast or slow AC mode (characterised by a real frequency $\omega _a=\omega_{k,l}$), and a stable MS wave (characterised by a real frequency $\omega _s$). The condition for the occurrence of near resonance interaction between those three eigenmodes is $\omega_s=\omega _a +\Delta \omega$, where the frequency mismatch $\Delta \omega$ is much smaller than each of the eigenfrequencies $\omega _a$ and $\omega _s$.
This condition is easily satisfied due to the continuous nature of the MS spectrum. The near-resonance condition can be satisfied also if the Gaussian axial mass distribution is considered. In that case the MS spectrum is discrete, however, as was pointed out in \cite{Shtemler:2014uq}, for any frequency $\omega _{k,l}$ of a stable AC mode, a corresponding eigenfrequency of the Gaussian MS spectrum may be found such that the near-resonance condition is satisfied.
The customary near-resonance condition on the axial  wave number is not needed here due to the axial stratification of the mass density; it is replaced by the solvability conditions of the higher orders boundary value problem in the axial coordinate.   The three-wave interaction is a direct result of the influence of the perturbed in-plain magnetic pressure gradients on the acoustic modes, and the simultaneous axial convection of the AC modes by the acoustic perturbations.

The main goal of the current section is to derive a set of coupled ordinary differential equations that govern the evolution of the amplitudes of the modes that take part in the near-resonance interaction. During the linear stage those amplitudes are constants that are determined by the initial conditions. However as the near-resonance interaction gains in importance, the mutual interaction changes dramatically their time evolution. Thus, while the equations that govern the amplitudes of the stable AC and MS modes are similar to their first order classical counterparts \cite{SagdeevandGaleev},  the equation for the amplitude of the MRI is of second order, reflecting the multiplicity two of the eigenvalue at the threshold beta (\cite{Shtemler:2013qv}, \cite{Stefani:2005fk}).

Deriving the amplitude equations starts with observing that for small values of the growth rate $\gamma$ the near-resonance interactions are described by two distinct time scales:  a fast time ${\bar \tau}=\omega _a t$ , and a slow time ${\tilde \tau}=\gamma t$. Assuming further that the frequency mismatch $\Delta \omega$ is of order $\gamma$, each of the perturbations due to the AC modes ($f(\eta, t)$) as well as those due to the MS waves ($g(\eta, t)$) may be represented as a sum of zeroth and first harmonic terms in ${\bar \tau}$ (higher harmonics are neglected), each with a slowly varying amplitude that depends on ${\tilde \tau}$:
\begin{equation}
f(\eta ,t)=f_0(\eta ,{\tilde \tau})+[f_1(\eta ,{\tilde \tau})e^{i{\bar \tau}}+c.c],
\label{AC}
\end{equation}
where the subscripts ($0$ or $1$) denote the harmonic number, and a similar expression for $g(\eta, t)$. Contributions to the zeroth harmonic AC perturbations come from the MRI, a non-resonant excitation of a zeroth harmonic MS wave, and the interaction of the stable fast AC and the MS eigenmodes, all of which may be represented respectively as:
\begin{eqnarray}
\nonumber
f_0(\eta ,{\tilde \tau})&=&A_0({\tilde \tau})P_0(\eta)+A_0({\tilde \tau})H_0({\tilde \tau})\psi_{0,0}(\eta)\\
&+&[A_1^*({\tilde \tau})H_1({\tilde \tau})\psi_{-1,1}(\eta)+c.c],
\label{zeroharmonic}
\end{eqnarray}
where $A_0$ is the real-valued amplitude of the MRI ($e^{\pm \tilde \tau}$ during the linear stage), $P_0$ its linear eigenfunction, $A_1, H_1$ are the amplitudes of the stable AC and MS modes, respectively (constants during the linear stage), $H_0$ is the amplitude of the non-resonantly excited MS wave, and $\psi_{-1,1}, \psi_{0,0}$ are yet to be determined coupling functions. In a similar manner the expressions for $f_1$ and $g_1$ are given by:
\begin{eqnarray}
f_1(\eta ,{\tilde \tau})=A_1({\tilde \tau})P_1(\eta)
+[A_0({\tilde \tau})H_1({\tilde \tau})\psi_{0,1}(\eta)+c.c]\label{firstharmonica}\\
g_1(\eta ,{\tilde \tau})=H_1({\tilde \tau})Q_1(\eta)
+[A_0({\tilde \tau})A_1({\tilde \tau})\phi_{0,1}(\eta)+c.c],
\label{firstharmonicm}
\end{eqnarray}
where $P_1, Q_1$ are the eigenfunctions of the AC and MS modes, respectively, $\phi _{0,1}(\psi _{0,1})$ is the coupling function between the MRI and the AC(MS) mode, and the first terms on the right hand sides of eqs. (\ref{zeroharmonic})-(\ref{firstharmonicm}) describe the three linear modes that participate in the near-resonance interaction. The amplitude of the non resonantly driven MS wave may be shown to be $H_0({\tilde \tau})=A_0^2({\tilde \tau})$ \cite{Shtemler:2014uq}.

The amplitudes $A_0, A_1$, and $H_1$ are the main players in the current work and deriving the set of ordinary differential equations that govern their slow-time dynamical evolution is the main concern of this section. Thus, recalling the second order degeneracy of the MRI at the threshold, the dynamical equation for $A_0$ is necessarily of second order, that in view of its near resonance interaction with the two other modes is conjectured to be given by:
\begin{equation}
\gamma ^2\frac{d^2 A_0}{d{\tilde \tau}^2}=\gamma ^2 A_0+\Gamma _{0,0}A_0H_0+[\Gamma _{-1,1}A_1^*H_1e^{-i{\theta\tilde \tau} }+c.c],
\label{amplitude}
\end{equation}
where $\gamma \theta = \Delta \omega$, and $\Gamma _{i,j}$ ($i,j$ indicate the pair of interacting harmonics) are coupling constants that measure the interaction between the various waves. A clear hierarchy emerges from eq. (\ref{amplitude}) according to which $A_0$ is of order $\gamma$ while $H_0, A_1$, and $H_1$ are of orders
$\gamma ^2, \gamma ^{3/2}$, and $\gamma ^{3/2}$, respectively. The corresponding equations for $A_1$ and $H_1$ are of first order and are identical with their classical counterparts (\cite{SagdeevandGaleev}, \cite{Shtemler:2013qv}):
\begin{equation}
\gamma \frac{dA_1}{d{\tilde \tau}}=i\Gamma _{0,1}A_0H_1e^{-i\theta {\tilde \tau}}, \; \gamma \frac{dH_1}{d{\tilde \tau}}=i\Gamma _{1,0} A_0A_1e^{i\theta {\tilde \tau}},
\label{firstorder}
\end{equation}
where $\Gamma _{i,j}$ are the corresponding coupling coefficients yet to be determined.

The calculation of the four real coupling coefficients ($\Gamma _{0,0}, \Gamma _{-1,1},  \Gamma _{0,1}, \Gamma _{1,0}$) is carried out by inserting eqs. (\ref{AC})-(\ref{firstorder}) into the MHD equations, which are subsequently solved order by order in $\gamma$, according to the hierarchy specified above. As expected, the lowest order reproduces the linear results. The next order yields four non homogeneous ordinary differential equations for the coupling functions ($\psi_{0,0}(\eta), \psi_{-1,1}(\eta), \psi_{0,1}(\eta), \phi_{0,1}(\eta)$). The four solvability conditions for those equations provide the four values for the sought  coupling coefficients. Of particular interest is the value of $\Gamma _{0,0}$ which is negative, a fact that plays a crucial role in the analysis of the resulting equations   \cite{Liverts:2012fr}.

Finally, by appropriately rescaling the amplitudes eqs. (\ref{amplitude})-(\ref{firstorder}) may be recast in the following form:
\begin{eqnarray}
\frac{d^2 a_0}{d{\tilde \tau}^2}=a_0-a_0^3+\sigma _0 \alpha ^2[a_1^*h_1e^{-i\theta {\tilde \tau}}+c.c]\label{rescaleda}\\
\frac{da_1}{d{\tilde \tau}}=i\alpha a_0h_1e^{-i\theta {\tilde \tau}}, \; \frac{dh_1}{d{\tilde \tau}}=i\sigma _1\alpha a_0a_1e^{i\theta {\tilde \tau}},
\label{rescaledah}
\end{eqnarray}
where $a_0, a_1, h_1$ are the rescaled amplitudes of the MRI, AC and MS modes, respectively,  $\alpha = \sqrt{|\Gamma _{0,1}\Gamma _{1,0}/\Gamma _{0,0}|}$, $\sigma _0 =sign(\Gamma _{0,1}\Gamma _{-1,1})$, and $\sigma _1 =sign(\Gamma _{0,1}\Gamma _{1,0})$.
Equations (\ref{rescaleda})-(\ref{rescaledah}) constitute the dynamical system that is investigated in the next section. Each interacting triad is thus characterised by a set of three coefficients $\alpha, \sigma _0$ and $\sigma _1$ which determines the nature to its dynamical evolution.

\textit{The Manley-Rowe relations} --
A convenient representation of the amplitudes of the AC and MS modes is given by
$a_1({\tilde \tau})=\rho _a({\tilde \tau})e^{-i\varphi _a({\tilde \tau})}, h_1({\tilde \tau})=\rho _h({\tilde \tau})e^{-i\varphi _h({\tilde \tau})}$
(recall that $a_0$ is real-valued). In terms of these variables
equations (\ref{rescaleda})-(\ref{rescaledah}) give rise to the following two constants of motion:
\begin{eqnarray}
E&=&\frac{1}{2}\Bigl (\frac{da_0}{d{\tilde \tau}}\Bigr )^2-\frac{1}{2}\Bigl (a_0^2-\frac{1}{2}a_0^4\Bigr )+F \label{manley1},\\
{\cal E}&=&\frac{1}{2}\rho _h^2 +\sigma _1\frac{1}{2}\rho _a^2 ,
\label{manley2}
\end{eqnarray}
where $F=\frac{1}{2}\sigma _0\alpha ^2\theta(\sigma _1\rho _h^2-\rho _a^2)-2\sigma _0a_0\alpha ^2\rho _a\rho _h cos\varphi$, \\*
$\varphi =\varphi _a-\varphi _h -\theta {\tilde \tau}$, and $E$ and ${\cal E}$ are constants.

\textit{Solutions of the dynamical amplitude equations} --
The presence of a frequency mismatch in the classical set of resonant three-wave interactions equations is known to lead to complex dynamical behaviour that include bifurcations to increasingly complicated periodic motions as well as apparent chaotic motions \cite{WERSINGER:1980fh}. Indeed, the solutions of eqs.
(\ref{rescaleda})-(\ref{rescaledah}) exhibit much more complex behaviour than their strict-resonance counterparts. As is demonstrated in \cite{Shtemler:2013qv} and \cite{Shtemler:2014uq} resonant triads with $\theta=0$ are divided into stable ($\sigma _1=-1$) and unstable ($\sigma _1=+1$) triads. In the unstable case the MRI mode saturates by nonlinearly coupling to AC and MS modes that grow exponentially as $e^{\gamma _{nl}\alpha \tilde{\tau}}$ with an effective growth rate $\gamma _{nl} \sim 1$, while the amplitudes within stable triads are bounded and exhibit oscillatory behaviour. This classification is carried over also to the near-resonant case and gives rise to the following two types of dynamical behaviour:
\\
\emph {i. Unstable triads - explosive instability}: Solving eqs. (\ref{rescaleda})-(\ref{rescaledah}) numerically indicate that unlike the strict-resonance case, all three modes may grow explosively in time, namely,  the solutions tend to infinity within a finite time, which is termed the explosion time and denoted by ${\tilde \tau}_e$. To simplify the calculations, the particular case $\rho _a({\tilde \tau})=\rho _a({\tilde \tau})\equiv \rho ({\tilde \tau})$ is considered. Numerical solutions demonstrate that for such triads, as
${\tilde \tau} \rightarrow {\tilde \tau}_e$ ,  $a_0$ and
$ \rho \rightarrow \infty$, while $\varphi \rightarrow (2m+1)\pi/2+\varphi _e$ where $m$ is an integer and $\varphi _e \rightarrow 0$,  such that $a_0\varphi _e$ remains finite. These observations lead to the ansatz $a_0({\tilde \tau})\varphi _e({\tilde \tau})=Const$ that results in the following asymptotic \textit{universal} expressions for the normalised amplitudes close to the explosion time:
\begin{equation}
{\hat a}_0\simeq \frac{K_0}{1-{\hat \tau}},\;\;\;\; {\hat \rho}\simeq \frac{K_1}{(1-{\hat \tau})^2}, \;\;\;\; {\hat \varphi} \simeq K_2(1-{\hat \tau}),
\label{explosive}
\end{equation}
where ${\hat \tau}\equiv {\tilde \tau}/{\tilde \tau}_e$, $K_0=-2(-1)^m$, $K_1^2=sign(\theta)6(1+4/\alpha ^2)$, and $K_2=-1/3$, and ${\hat a}_0, {\hat \rho}, {\hat \varphi}$ are the appropriately normalised amplitudes and phase, respectively. The explosion time cannot be calculated by the analysis that leads to eq. (\ref{explosive}) and is
determined by the initial conditions. It is evident that under the ansatz specified above explosive solutions exist only for $\theta >0$. Numerical solutions demonstrate however that such solutions exist also for $\theta <0$, for which $a_0({\tilde \tau})\varphi _e({\tilde \tau})$ is not a constant near the explosion time.

Figure \ref{fig1} presents a comparison between the numerical solution of eqs. (\ref{rescaleda})-(\ref{rescaledah}) and the asymptotic solutions (\ref{explosive}). The triad is composed of an MRI mode ($k=1, l=-1$), a stable slow AC mode ($k=2, l=-1$), and a stable MS mode. For that triad $\sigma _1=-1$ and $\alpha =20.5$.
It is seen that after an initial transient period, as time approaches the explosive time, the two dashed lines, which represent the numerical solutions of eqs. (\ref{rescaleda})-(\ref{rescaledah}) for two different values of ${\hat \theta}=\theta /\alpha$, converge to the same full line, which represents the universal asymptotic solution given by eq. (\ref{explosive}). The explosion time as obtained from the full numerical solutions of eqs.  is $\alpha \tilde{\tau}_e \sim 4.5$.  The triad depicted in Figure \ref{fig1} is characterised by the biggest value of $\alpha$ as compared to all triads with slow AC modes, and hence has the shortest explosion time. In this sense, this first unstable triad is deemed the most unstable.  It should be finally mentioned that it is the negative sign of $\sigma _1$ that enables the unbounded
growth of the amplitudes of the participating modes while keeping ${\cal E}$ constant according to the Manley-Rowe relation (\ref{manley2}).\\

\begin{figure}[h]
\centering
\includegraphics*[width=85mm,height=35mm]{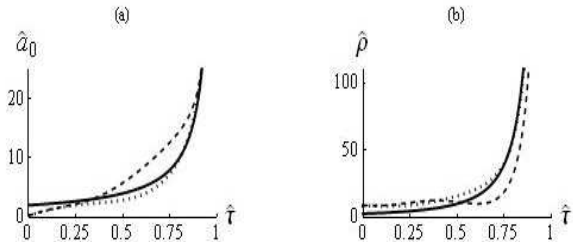}
\caption{Comparison of the numerical solution of eqs. (\ref{rescaleda})-(\ref{rescaledah}) and the universal asymptotic solution (\ref{explosive}) for a triad characterised by $\sigma _1=-1, \alpha =20.5$. Full line: eq. \ref{explosive}, upper (lower) dashed line: solution of eqs. (\ref{rescaleda})-(\ref{rescaledah}) with $\hat{\theta}=2.5\;(1)$.}
\label{fig1}
\end{figure}

\noindent \emph {ii. Stable triads - irregular oscillations}:
The amplitudes of the three modes that form a stable triad remain bounded at all times and exhibit irregular periodic oscillations. This may be seen in Fig. \ref{fig2} where in particular saturation of the MRI is demonstrated.

\begin{figure}[h]
\centering
\includegraphics*[scale=1.25]{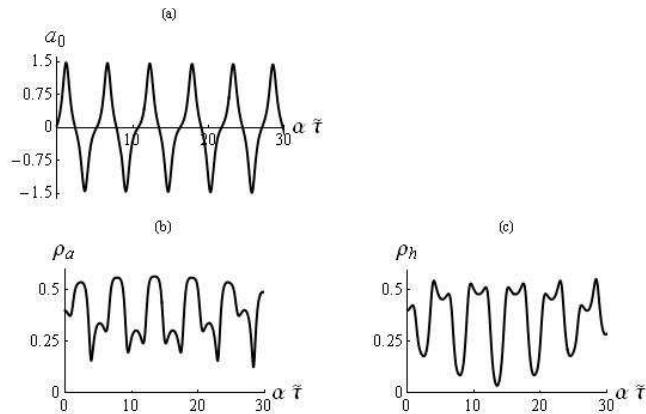}
\caption{Solution of eqs. (\ref{rescaleda})-(\ref{rescaledah}) for a stable triad that is composed of an MRI, a stable fast AC mode ($k=1,l=1$) and an MS mode.}
\label{fig2}
\end{figure}

\textit{Conclusions - } The nonlinear evolution of near-resonance triads of modes (composed of MRI, and stable AC and MS modes) is studied for thin Keplerian disks. The slowly varying amplitude of the MRI is governed by a Duffing equation with a force term that is time-dependent due to the frequency mismatch, while the amplitudes of the other two modes are determined by a first order rate equation. It is shown that if the stable AC mode is a slow one  all the amplitudes grow explosively in time while in triads that include a stable fast AC mode all amplitudes are bounded and exhibit irregular oscillations. It is conjectured that the explosive instability describes an intermediate stage of the nonlinear evolution of the MRI during which a significant energy transfer between the modes take place, before the blow-up and consequent break-down of the present model. Direct numerical simulations can resolve the evolution beyond the explosion time.

%
%
\textit{Acknowledgment -} This work was supported by grant number 180/10 of the Israel Science Foundation.

\end{document}